\documentclass[twocolumn]{aastex631}
\shorttitle{
Radial drift in warped protoplanetary disks
}
\usepackage{amsmath}	% Advanced maths commands
\usepackage{amssymb}	% Extra maths symbols
\usepackage{float}
\usepackage{ulem}
\usepackage{soul}
\usepackage{lipsum}
\usepackage[T1]{fontenc}
\usepackage{rotating}

\DeclareRobustCommand{\VAN}[3]{#2}
\let\VANthebibliography\thebibliography
\def\thebibliography{\DeclareRobustCommand{\VAN}[3]{##3}\VANthebibliography}

\begin{document}

%% LaTeX will automatically break titles if they run longer than
%% one line. However, you may use \\ to force a line break if
%% you desire.

\title
{
Radial drift in warped protoplanetary disks 
}
\shortauthors{Rozner \& Perets}
%% Use \author, \affil, and the \and command to format
%% author and affiliation information.
%% Note that \email has replaced the old \authoremail command
%% from AASTeX v4.0. You can use \email to mark an email address
%% anywhere in the paper, not just in the front matter.
%% As in the title, use \\ to force line breaks.

%\author[0000-0002-2728-0132]{Mor Rozner}
%\affiliation{Technion - Israel Institute of Technology, Haifa, 3200002, Israel}

%\collaboration{6}{(AAS Journals Data Editors)}
\email{morozner@campus.technion.ac.il}

\author[0000-0002-2728-0132]{Mor Rozner}
\affiliation{Technion - Israel Institute of Technology, Haifa, 3200002, Israel}

\begin{abstract}
The meter-size barrier in protoplanetary disks is a major challenge in planet formation, for which many solutions were suggested. One of the leading solutions is dust traps, that halt or slow the inward migration of dust particles.
The source and profile of these traps are still not completely known.
Warped disks are ubiquitous among accretion disks in general and protoplanetary disks in particular, and the warping could lead naturally to the formation of dust traps.
Dust traps in warped disks could rise not only from pressure gradients, but also due to different precession rates between gas and dust. 
Here we derive analytically the radial drift in warped disks, and demonstrate derivation for some specific conditions. The radial drift in warped protoplanetary disks is qualitatively different, and depending on the structure of the disk, dust traps could form due to the warping. 
Similar processes could lead to the formation of traps also in other accretion disks such as AGN disks. 
\end{abstract}
% Select between one and six entries from the list of approved keywords.

%%%%%%%%%%%%%%%%%%%%%%%%%%%%%%%%%%%%%%%%%%%%%%%%%%

%%%%%%%%%%%%%%%%% BODY OF PAPER %%%%%%%%%%%%%%%%%%
\section{Introduction}

Protoplanetary disks do not form and evolve in isolation.
Most stars are formed in groups or stellar clusters \citep{LadaLada2003}, in which many stars are born and evolve. Planets form in disks of gas and dust surrounding these stars, such that the environment of planet formation is far from being quiescent, and the interactions with the environment could play a crucial role in shaping the conditions and properties of planet formation.
Protoplanetary disks could be affected by flybys, binarity of the host star and magnetic interactions with the protostar (e.g. \citealp{FoucartLai2011,LubowMartin2018,Nealon2018,Kraus2020}). 
All these interactions leave signatures on the disk, such as disk warping and even breaking and extreme cases.
And indeed, there is governing observational evidence for distorted protoplanetary disks (e.g. \citealp{Benisty2015,Benisty2018,Bi2020GWOri}), that indicates the importance of modeling planet formation in them. However, the majority of planet formation theories consider flat, isolated disks. 

The growth of small dust grains to full-sized planets is assumed to be hierarchical, i.e. from dust grains to pebbles, planetesimals and so on, and includes several orders of magnitude. 
The initial stages, up to cm size, as well as the final stages, from km-size to planetestimals, are well explained by the current planet formation theories -- by sticking \citep{WurmBlum1998} and pebble accretion \citep{OrmelKlahr2010,PeretsMurrayClay2011,Lambrechts2012}, correspondingly. However, meter-size objects have to overcome several barriers in order to enhance their growth, and although many solutions were suggested to this problem, it is still an open question and a subject of ongoing research (see review \citealp{Morbidelli2016}).
One of the major barriers in planet formation is the radial-drift barrier, which describes the rapid migration of objects toward the star, due to the pressure gradient in the disk, in shorter timescales than the expected growth timescales  \citep{Weidenschilling1977}.  
Further barriers include collisional fragmentation, erosion (e.g. \citealp{Guttler2010}) and aeolian-erosion (e.g. \citealp{Demirci2020,Erosion1,Erosion2}).
There are several suggested solutions to the meter-size barriers, we will list some of the major ones. 

Streaming-instability is a promising suggested solution to overcome the meter-size barriers \citep{YoudinGoodman2005}, in which the local concentration of solids in the disk is catalyzed to a point where gravitational collapse could take place and give rise to the formation of larger objects. 
Another suggested solution is the seeding of already-formed objects, including km-size objects. In this approach, low-probability events of planet formation are sufficient to explain the total number of exoplanets, due to an external transport of planet-seeds, such that together with streaming-instability for example, the whole population of planets could be explained \citep{GrishinPeretsAvni2019}.
Another suggested solution is dust traps. The first dust traps to be suggested were pressure traps, in which local pressure maxima in the disk induce zero pressure gradient, which suppresses the usual radial drift and leads to a dust pile-up\citep{Nakagawa1986,Whipple1972}. Several mechanisms were suggested as sources for these traps, including vortices (e.g. \citealp{BargeSommeria1995}), planet gap edges (e.g. \citealp{Pinilla2012gap_edges}) and self-induced \citep{Gonzalez2017}. Recently, it was suggested that a dust trap could be formed not only by pressure gradient, but also due to precession difference \cite{AlyLodato2020,Longarini2021}. In distorted disks, the gas and dust precess differently, such that dust rings are formed, and there are locations in which the relative velocity between gas and dust becomes zero and the radial drift is suppressed. 

In this paper, we extend and generalize the analytical study of dust traps in warped disks and the radial drift there in general. We derive the modified equations of motion for dust and gas particles in warped disks, and present the steady-state solutions in a closed analytic form under some conditions. By that, we generalize the standard radial drift expression used in flat disks, and give a complementary perspective to the hydrodynamical works on warped disks mentioned earlier. 
Our derivation is general and should apply to any warped disk, regardless of the warping source. We manifest the derivation for typical parameters of warped circumbinary disks and discuss implications for these disks and other.

The usual radial drift equations are derived usually based on the assumptions of a flat, axisymmetric disk (e.g. \citealp{Nakagawa1986}), here we relieve some of the initial assumptions to describe analytically the radial drift in warped disks and discuss the implications of the modified radial drift. Due to the assymetry of warped disks, dust traps could form not only due to the existence of pressure gradients but also due to the structure of the disk, which will lead to another kind of dust traps, distinguished from the pressure traps by its nature.

In section \ref{sec:disk structure} we describe the structure of a warped disk. In section \ref{sec:gas and dust evolution} we describe the coupled evolution of gas and dust and the unique behaviour in warped disks. In section
we present the modifications for the standard radial drift in the case of circumbinary warped disks \ref{sec:circumbinary}.
In section
\ref{sec:discussion and implications} we discuss the radial drift in several additional examples of warped disks and discuss further implications. In section \ref{sec:caveats} we discuss the caveats of our model. In section \ref{sec:summary} we summarize the paper and conclude. 

\section{The structure of a warped disk}\label{sec:disk structure}

%\begin{figure}[H]
 % \includegraphics[width=1\linewidth]
%{graphs/disk figure}
 % \caption{An illustration of a warped circumbinary disk.
  %}
  %\label{fig:illustration}
%\end{figure}

%Consider a circumbinary disk surrounding a binary with masses $m_1$ and $m_2$, a total binary mass of $m_{\rm bin}=m_1+m_2$ and separation $a_{\rm bin}$ in a circular orbit. The disk surface density is $\Sigma_g(r)$
Consider a disk that
extends from $r_{\rm in}$ to $r_{\rm out}$.
The disk density profile and thickness are given by 

\begin{align}
\Sigma_g(r) \propto r^{-p}, \
\frac{H}{r}\propto r^{(2p-1)/4}
\end{align}

\noindent
where we use $p=3/2$ following \cite{Armitage2011}.
The orientation of the disk is specified by the normalized angular momentum vector $\hat{\textbf{l}}(r)$, 

\begin{align}
\hat{\textbf{l}}(r) = (\sin \beta\cos \gamma, \sin \beta \sin \gamma, \cos \beta)
\end{align}

\noindent
where $\beta(r)$ is the warp angle and $\gamma(r)$ is the twist angle. At $r=r_{\rm in},r_{\rm out}$, the angular momentum directions are $\hat{\textbf{l}}_{\rm in},\hat{\textbf{l}}_{\rm out}$ correspondingly. A disk is defined as warped if $\beta(r)$ varies with the radius.
%The angular momentum direction of the binary is $\hat{\textbf{l}}_{\rm b}$. %See Fig. \ref{fig:illustration} for an illustration not for scale. 

There are two regimes of warps propagation: the diffusive regime ($\alpha>h/r$) and the wavelike regime ($\alpha<h/r$) in which the warps propagate as bending waves and pressure forces drive the evolution. $\alpha$ is the Shakura Sunayev parameter \citep{ShakuraSunyaev1973} and $h/r$ is the aspect ratio of the gas. Warped protoplanetary disks are described by the wavelike regime (e.g. \cite{PapaloizouLin95,LubowOgilvie2000} and references therein), while AGN disks for example are in the diffusive regime (e.g. \citealp{PapaloizouTerquemLin1998,LodatoPrice2010}).    
The steepness of the warp could be described by the warp amplitude  dimensionless radial change in the angular momentum, $\psi := r |\partial \hat{\bf l}/\partial r|$. When $\psi/(h/r)\gg 1$, the bending waves become non-linear.

\section{Gas \& Dust Evolution}\label{subsec: dust evolution}\label{sec:gas and dust evolution}

Objects in protoplanetary disks experience gas drag, which depends on the size of the object, its velocity relative to the gas and the gas properties. The overall gas drag in the different regimes could be written as (e.g. \citealp{PeretsMurrayClay2011})  

\begin{align}
\textbf F_D = - \frac{1}{2}C_D(Re) \pi R^2 \rho_g v_{\rm rel}^2\hat{\textbf{v}}_{\rm rel}
\end{align}

\noindent
such that $R$ is the size of the object, $\rho_g$ is the gas density, $v_{\rm rel}$ is the relative velocity between the object and the gas and $C_D(Re)$ is a function that depends on the Reynolds number $Re=2Rv_{\rm rel}/(0.5 v_{\rm th}\lambda)$. $v_{\rm th}=\sqrt{8/\pi}c_s$ is the thermal velocity, $c_s$ is the sound speed and $\lambda$ is the mean free path of the gas. For small objects, the force scales linearly with the relative velocity, i.e. $\textbf F_{\rm D}\propto -\textbf{v}_{\rm rel}$, where for large objects the force scales quadratically, i.e. $\textbf{F}_{\rm D}\propto -v_{\rm rel}^2\hat{\textbf{v}}_{\rm rel}$.

The dynamics of dust is affected by the interaction with the gas.
While small dust grains are well-coupled to the gas, larger grains acquire relative velocity.
The coupling to the gas is encapsulated by the Stokes number, defined by 

\begin{align}
St = \Omega_K t_{\rm stop}, \ 
t_{\rm stop}=\frac{|mv_{\rm rel}|}{|F_D|}
\end{align}

\noindent

where $\Omega_K$ is the Keplerian orbital frequency, $t_{\rm stop}$ is the stopping time, $v_{\rm rel}$ is the relative velocity and $F_D$ is the drag force. The dependence of $t_{\rm stop}$ on the size of the grain, $R$, varies with the corresponding regime, i.e. Epstein ($R \lesssim \lambda_{\rm mfp}$), Stokes ($R\gtrsim \lambda_{\rm mfp}, \ Re<1$ where $Re$ is the Reynolds number) or ram-pressure ($Re\gtrsim 800$) where $\lambda_{\rm mfp}$ is the mean free path of the gas. For the Epstein and Stokes regimes (e.g. \citealp{PeretsMurrayClay2011}),  

\begin{align}
t_{\rm stop}=
\begin{cases}
\frac{\rho_p}{\rho_g}\frac{R}{ v_{\rm th}},\ \ $Epstein$, \\
\frac{4}{9}\frac{\rho_p}{\rho_g}\frac{R^2}{\lambda_{\rm mfp} v_{\rm th}}, \ \ $Stokes$
\end{cases}
\end{align}

\noindent
where $\rho_g$ is the gas density, $\rho_p$ is the object's density and $v_{\rm th}=\sqrt{8k_BT/(\pi \mu m_H)}$ is the thermal velocity, $T$ is the temperature of the gas and $\mu m_H$ is the mean molecular weight. 

In this paper, we focus on the regime $R\gtrsim \lambda$, in which the drag force on the dust is given by (\citealp{Nakagawa1986} and references therein),
\begin{align}
&\textbf{F}_{D}= -A\rho_g (\textbf{v}_d-\textbf{v}_g), \\
%&A = \frac{4}{3}\pi \rho_g \bar v_{\rm th}R^2
&A = \frac{3\bar v_{\rm th}\lambda}{2\rho_s R^2} 
\end{align}
\noindent
where similar equations could be written for the gas, replacing $\rho_g$ with $\rho_d$. $R$ is the size of the object. Unless stated otherwise, the fiducial parameters we consider are $\rho_g=3\times 10^{-9}\left(a/a_0\right)^{-16/7 } \ \rm{g \ \rm{cm}^{-3}}, \ \lambda = 1 \left(a/a_0\right)^{16/7} \ \rm{AU}, \ \bar v_{\rm th}=\sqrt{8k_BT/\mu m_H \pi}
, \ T=120 \ \left(a/a_0\right)^{-3/7} \ \rm{K}, \ \mu m_H=3.9\times 10^{-24} \ \rm{g}, \ \rho_s= 1 \ \rm{g \ cm^{-3}}$ (\citealp{Nakagawa1986,PeretsMurrayClay2011} and references therein), 
where $a_0$ is a characteristic lengthscale, and we focus on meter-size objects, i.e. $R= 1 \ \rm m$. The dust-to-gas ratio is taken to be $1\%$ \citep{ChiangYoudin2010}.

The steady-state radial drift of the dust, when effects of backreaction are neglected, is given by (e.g. \citealp{Nakagawa1986,Birnstiel2016})

\begin{align}
&\frac{dr}{dt}=v_r=-\frac{\rho_g}{\rho_g+\rho_d}\frac{2D\Omega_K}{D^2+\Omega_K^{\textbf{2}}}\eta v_K, \label{eq: flat radial drift} \\
&\eta = -\frac{1}{2}\frac{1}{\rho_g}\frac{\partial P_g}{\partial r}v_K\Omega_K
\end{align}

\noindent
where $v_K$ is the Keplerian velocity, $\Omega_K$ is the Keplerian angular velocity, $\bf{D=A(\rho_g+\rho_d)}$,
$P_g$  is the gas pressure and $\eta$ is the gas pressure support parameter.

The gas and dust velocity fields could be thought as a rotation of the velocity fields in a flat Cartesian disk (see also \cite{Longarini2021}). 
\begin{widetext}
\begin{align}\label{eq:u decomposition main text}
\textbf{u}_d=\begin{pmatrix}
u'_{d,r}\left(\cos\phi\cos \beta\cos \gamma_d-\sin \gamma_d\sin \phi\right)-u'_{d,\phi}\left(\sin \phi \cos \beta \cos \gamma_d+\cos \phi \sin \gamma_d\right) \\
u'_{d,r}\left(\cos \phi \sin \gamma_d \cos \beta +\cos \gamma_d\sin \phi\right)+u'_{d,\phi}\left(\cos \phi\cos \gamma_d-\sin \phi \sin \gamma_d\cos \beta\right)\\
-u'_{d,r}\sin \beta\cos \phi +u'_{d,\phi}\sin \phi \sin \beta
\end{pmatrix},\\~\\
\textbf{u}_g=\begin{pmatrix}
u'_{g,r}\left(\cos\phi\cos \beta\cos \gamma_g-\sin \gamma_g\sin \phi\right)-u'_{g,\phi}\left(\sin \phi \cos \beta \cos \gamma_g+\cos \phi \sin \gamma_g\right) \\
u'_{g,r}\left(\cos \phi \sin \gamma_g \cos \beta +\cos \gamma_g\sin \phi\right)+u'_{g,\phi}\left(\cos \phi\cos \gamma_g-\sin \phi \sin \gamma_g\cos \beta\right)\\
-u'_{g,r}\sin \beta\cos \phi +u'_{g,\phi}\sin \phi \sin \beta
\end{pmatrix}
\end{align}
\end{widetext}

\noindent
where we consider $\beta=\beta_{g}=\beta_d$, $\Omega_p$ is the rigid precession of the gas such that $\gamma_g = \Omega_p t$ and $\Omega_{\rm ext}$ is the dust precession, $\gamma_d = \Omega_{\rm ext}t$ and $\phi$ is the polar angle along the disk. 
The velocity field in a warped disk is then a rotation of the velocity in a flat disk by angle $\beta$ around the y-axis and $\gamma_i$ around the z-axis. 
The velocity components in the flat coordinate system are given by $u'_{i,j}$, where subindex of d relates to dust components, g to gas components and $r$ and $\phi$ to the spatial component in a cylindrical coordinate system. For simplicity, we ignore the vertical component in the flat system, i.e. we consider $u'_{d,z}=u'_{g,z}=0$. 

The gas and the dust precess in different frequencies, $\Omega_p$ and $\Omega_{\rm ext}$, where the gas precesses rigidly and the dust precess differentialy. These two frequencies become equal at the co-precession radius, in which a dust-ring is formed, as was shown in a hydrodynamical simulation in \cite{AlyLodato2020}. The location of the dust ring is not correlated with pressure maxima, such that the dust trap potentially formed is essentially different then the usual pressure traps discussed \citep{Aly2021,Longarini2021}. Here we derive analytically a full expression for the radial drift in warped disks.

The equations of motion of the gas and the dust are given by \citep{Nakagawa1986}

\begin{align}
&\frac{d\textbf u_d}{dt} =-A \rho_g(\textbf{u}_d-\textbf{u}_g)-\frac{GM_\star}{r_d^3}\textbf{r}_d, \\
&\frac{d\textbf u_g}{dt} =-A \rho_g(\textbf{u}_g-\textbf{u}_d)-\frac{GM_\star}{r_g^3}\textbf{r}_g-\frac{\nabla P_g}{\rho_g}
\end{align}

\noindent
where $A$ is the gas drag coefficient 
and these equations could be solved substituting the decomposition introduced in eq. \ref{eq:u decomposition main text}, see Appendix \ref{app: full derivation} for further details of the derivation. The radial drift in a flat disk (eq. \ref{eq: flat radial drift}) is then modified, and a full analytical solution could be derived. In the co-precession radius, in which the dust ring is formed, the modification of the radial drift could be given by 

\begin{align}
&v_{dr} = -\frac{\rho_g}{\rho_g+\rho_d}\frac{2D\Omega_K}{D^2+\Omega_K^{\textbf{2}}}\eta v_K \xi(\gamma,\beta), \label{eq: modified radial drift}\\
&\xi(\gamma,\beta) =\frac{(G_\phi+G_r)(D^2+\Omega_K^2)}{D\left[C_r(G_r+G_\phi)D-0.5C_\phi \Omega_K\right]+\Omega_K^2} 
\end{align}

\noindent
$G_\phi, \ G_r, \ C_r$ and $C_\phi$ are given explicitly in Appendix \ref{app: full derivation} and $\gamma:=\gamma_d=\gamma_g$, such that for $G_\phi=1, \ G_r=0, \ C_r=1$ and $C_\phi=0$ we retrieve the usual radial drift for a flat disk.  

As can be seen from eq. \ref{eq: modified radial drift}, in warped disks a dust trap could form not only due to pressure maxima ($\eta=0$) but also due to precession ($\xi=0$). Even if for some choices of parameters, the radial drift won't vanish completely, its profile differ significanly from the usual radial drift considered in flat disks.

The precession frequencies are determined by the structure of the disk and the external torques applied on the disk. 
Given an external torque with a corresponding external precession frequency $\Omega_{\rm ext}$, the external torque density is given by $\bf{T= \Omega_p \times L}$, where $\textbf{L}=\Sigma r^2 \Omega \textbf{l}$ is the angular momentum of the disk per unit area. $\Sigma$ is the surface density, $\Omega$ is the angular frequency and $\textbf l(r)$ is a unit vector in the local direction of the angular momentum. The global precession frequency of the disk $\Omega_p$ (rigid precession), is defined by $T_{\rm tot}=\Omega_p L_{\rm tot}$, where for convenience we set $\hat z:=\hat \Omega_{\rm ext}$, which enables us to move to a non-vector equation.  We will follow briefly the derivation in \cite{LodatoFacchini2013}. For small warps, the overall torque and angular momentum could be calculated by 

\begin{align}
&T_{\rm tot}\approx \int_{r_{\rm in}}^{r_{\rm out}}\Omega_{\rm ext}(r)L(r)2\pi r dr, \\
&L_{\rm tot}\approx \int_{r_{\rm in}}^{r_{\rm out}}L(r) 2\pi r dr
\end{align}

\noindent
such that $\Omega_p$ is given by 

\begin{align}
\Omega_p = \frac{\int_{r_{\rm in}}^{r_{\rm out}}\Omega_{\rm ext}(r)L(r)2\pi r dr}{\int_{r_{\rm in}}^{r_{\rm tot}}L(r) 2\pi r dr}
\end{align}

\noindent
Assuming $\Omega_{\rm ext}(r)\propto r^{-s}$ and $L(r)\approx \Sigma r^2 \Omega \sqrt{l_x^2+l_y^2}\propto r^{1/2-p}$, the expression could be simplified for  $0<p<5/2$

\begin{align}\label{eq: Omegap}
&\Omega_p=\Omega_{\rm ext} (r_{\rm in})\frac{\int_1^{r_{\rm out}/r_{\rm in}}x^{3/2-p-s}dx}{\int_1^{r_{\rm out}/r_{\rm in}}x^{3/2-p}dx}= \\ \nonumber
&=\Omega_{\rm ext}(r_{\rm in})\frac{1-\left(r_{\rm out}/r_{\rm in}\right)^{5/2-p-s}}{\left(r_{\rm out}/r_{\rm in}\right)^{5/2-p}-1}\frac{5/2-p}{s+p-5/2}
\end{align}

%\section{Discussion}\label{sec:discussion and implications}

%\sub
\section{Circumbinary/Circumtriple disks}\label{sec:circumbinary}

Radial drift in circumbinary disks is even faster than in single-star disks \citep{Zagaria2021}. In general, there were set severe constraints on the possibility of planet formation in circumbinary disks (e.g. \citealp{Thebault2006} and references therein). The lifetime of a circumbinary disk is shorter due to the massive depletion of solids, which leaves a narrower available parameter space for planets' growth. However, there are many observations of planets around binary systems (e.g. \citealp{Marzari2019}). These together give greater importance to understanding dust traps in binary disks, either pressure traps \citep{Nakagawa1986,Whipple1972} or traffic jams \citep{AlyLodato2020,Aly2021,Longarini2021}. 

Consider a circumbinary containing companions with masses $m_1$ and $m_2$ and a separation of $a_{\rm bin}$. The binary imposes a torque per unit area on the disk element at a given radius $r$, averaged over a binary orbital period and the disk azimuthal direction, is specified by \citep{FoucartLai2013}

\begin{align}
\textbf{T}_{\rm bin}(r)= -\frac{3Gm_1m_2\Sigma_ga_{\rm bin}^2}{4(m_1+m_2)r^3}(\hat{\textbf{l}}_b\cdot \hat{\textbf{l}})(\hat{\textbf{l}}_b\times \hat{\textbf{l}})
\end{align}

\noindent
where $\hat l_b$ is the direction of the angular momentum of the binary. Then, 

\begin{align}
&\Omega_{\rm ext}(r) = \frac{3 Gm_1m_2 a_{\rm bin}^2}{4(m_1+m_2)r^5 \Omega_{\rm bin}(r)}\propto r^{-7/2}, \\
&\Omega_{\rm bin}=\sqrt{\frac{G(m_1+m_2)}{r^3}} 
\end{align}

\noindent
and following eq. \ref{eq: Omegap}, 

\begin{align}
\Omega_p =\frac{2}{5}\Omega_{\rm ext}(r_{\rm in}) \frac{1-\left(r_{\rm out}/r_{\rm in}\right)^{-5/2}}{\left(r_{\rm out}/r_{\rm in}\right)-1}
\end{align}

\noindent
As discussed in \cite{AlyLodato2020,Aly2021,Longarini2021}, dust traps are expected to form in circumbinary disks and here we extended their model analytically. 

Consider a circumbinary disk with $m_1=m_2= 1 \ M_\odot$, and following \cite{Longarini2021}, $a_{\rm bin}= 10 \ \rm{AU}$, $\beta=\pi/6$, $R_{\rm in}= 15 \ \rm{AU}$ and $R_{\rm out}= 150 \ \rm{AU}$. 

\begin{figure}[ht!]
  \includegraphics[width=1\linewidth]
{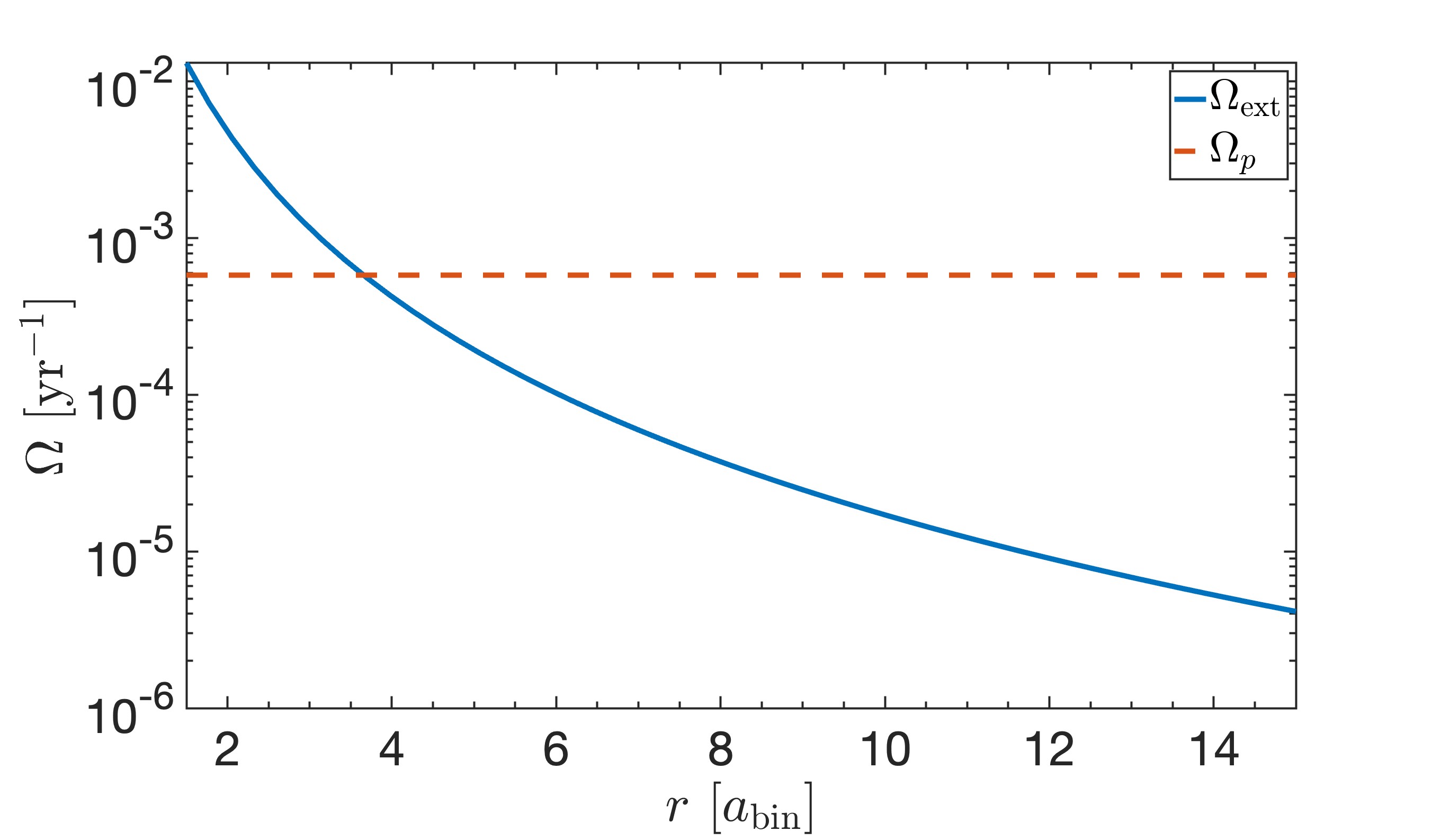}
  \caption{The differential precession frequency of the dusty disk $\Omega_{\rm ext}$ and the rigid precession frequency of the gaseous disk $\Omega_p$ for $m_1=m_2= 1 \ M_\odot, \ a_{\rm bin}= 10 \ \rm{AU}$, $R_{\rm in}= 15 \rm{AU}$ and $R_{\rm out}= 150 \rm{AU}$. 
  }
  \label{fig:Omegas}
\end{figure}

In Fig. \ref{fig:Omegas}, we present the differential precession frequency of the dusty disk $\Omega_{\rm ext}$ and the rigid precession frequency of the gaseous disk $\Omega_p$ for the parameters we specified earlier. As can be seen, for this choice of parameters, the co-precession radius is at $R_{co}\approx 3.7 a_{\rm bin}$. For these parameters, we will evaluate the correction to the radial drift at the co-precession radius.   
\\
We orbit-average numerically the coefficients introduced in Appendix \ref{app: full derivation}, to obtain the effective phase-independent $\xi$. 
\begin{figure}[ht!]
  \includegraphics[width=1\linewidth]
{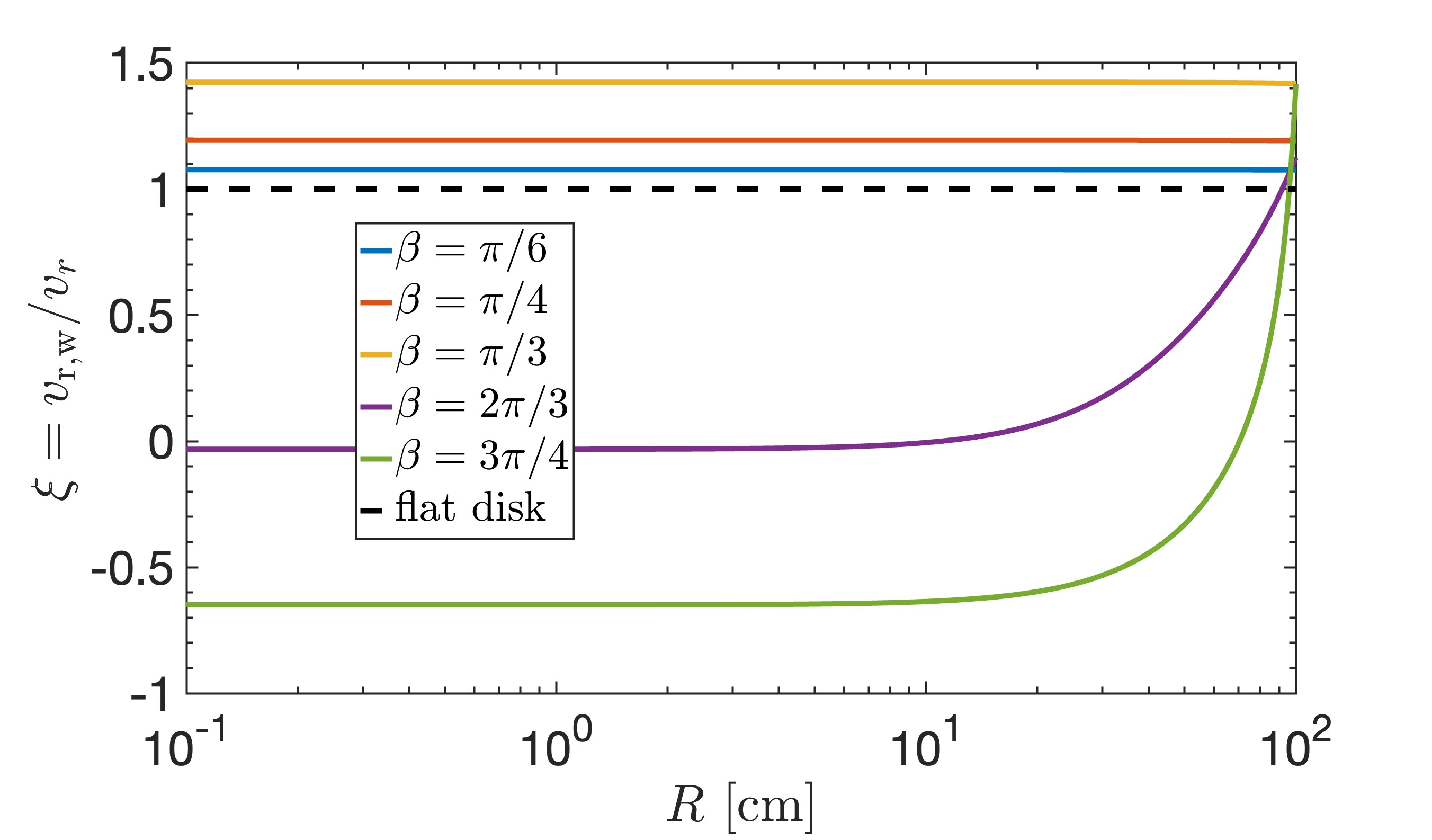}
  \caption{The correction to the standard radial drift in a warped circumbinary disk with  $m_1=m_2= 1 \ M_\odot, \ a_{\rm bin}= 10 \ \rm{AU}$, $R_{\rm in}= 15 \rm{AU}$ and $R_{\rm out}= 150 \rm{AU}$, for different tilt angles $\beta$, calculated in the co-precession radius. 
  }
  \label{fig:xi}
\end{figure}
In Fig. \ref{fig:xi}, we present the averaged correction factor $\xi$, for different choices of the tilt angles $\beta$. As can be seen, while for some tilt angels the radial drift for objects smaller than meter-size is larger than the standard radial drift derived in \cite{Weidenschilling1977}, for some other angles the correction due to warping could lead to a significantly suppressed radial drift and even radial drift in the opposite direction. 
Our results could explain also the enhanced radial drift found in the other numerical simulations of circumbinary disks (e.g. \citealp{Zagaria2021}).
As can be seen, the radial drift is essentially different in warped disks, whether it is suppressed or enhanced, and hence the initial stages of planet formation in these disks.

Recently, a misaligned circumtriple protoplanetary disk was found \citep{Bi2020GWOri,Smallwood2021}. The triple structure is more complicated and gives rise potentially to warps and breaks in the disk. Hence, enhanced formation of dust traps is expected also in circumtriple disks.

\section{Discussion}\label{sec:discussion and implications}

In the following, we will discuss radial drift in several examples of warped disks, distorted disks in general and implications. 

\subsection{Distant stellar companion}

Consider a central star with mass $M_1$, surrounded by a circumstellar disk. This system is in orbit with a distant binary companion of mass $M_2$, with a separation $a_{\rm bin}$. The binary companion exerts a perturbing torque on the disk. Averaging over the orbital period of the disk annulus and the binary, the torque per unit mass is given by \citep{ZanazziLai2018}

\begin{align}
&\textbf T_{\rm ext} = -r^2 \Omega \omega_{db} (\hat{\textbf{l}}_b\cdot \hat{\textbf{l}})(\hat{\textbf{l}}_b\times \hat{\textbf{l}}), \\ 
&\Omega(r) \approx \sqrt{\frac{GM_\star}{r^3}}, \
\omega_{\rm{db}} (r)=\frac{3GM_\star}{4a_{\rm bin}^3 \Omega}
\end{align}

For this case, $\Omega_{\rm ext}=\omega_{db}$ and $s=3/2$ in the notation of the previous subsection. Hence, the rigid precesseion is given by 

\begin{align}
\Omega_p = 2 \Omega_{\rm ext}(r_{\rm in})\frac{1-\sqrt{r_{\rm in}/r_{\rm out}}}{(r_{\rm out}/r_{\rm in})-1}
\end{align}

\noindent
The dust ring will then form where $\Omega_{\rm ext}=\Omega_p$, 

\begin{align}
r_{\rm dr}= 2^{2/3}r_{\rm in} \left(\frac{1-\sqrt{r_{\rm in}/r_{\rm out}}}{(r_{\rm out}/r_{\rm in})-1}\right)^{2/3}
\end{align}

Here, for some choices of parameters, the dust ring radius could be smaller than the inner radius, indicating that there would be no formation of a dust ring. However, the radial drift profile will still change according to the equations of motion of the gas and dust we introduced.

\subsection{Eccentric disks}

Eccentric disks are another class of distorted disks, and as such, the radial drift in them might be modified as well. The eccentricity of the disk could be perturbed again by an external torque or planet embedded in the disk. In principle, the radial drift on these disks could be derived simlarly to the derivation we introduced in Appendix \ref{app: full derivation}, but with different coefficients for the equation.

\subsection{Flybys}

 While binary interactions could maintain long-living significant warps, flybys could lead only to more moderate effects \citep{Nealon2020}.
 The effect of flybys on planet formation is expected to be more limited, since the warp-damping timescale in this case could be shorter than the typical growth timescale of a planet. Even in this case, pebbles might be able to grow to a large enough size to reduce the effect of the radial drift. Moreover, several flybys could have a cumulative effect on the planet formation in the disk.
 
\subsection{Symbiotic relations with other processes in the disk}

Perturbations in the disk such as warps also affect other growth/destruction processes in planet formation apart from the formation of pressure bumps. Once the dust-to-gas ratio in the vicinity of the trap is large enough, a runaway process will start and the growth will be enhanced even more, due to the formation of a dust trap \citep{Gonzalez2017}.
As we discussed in subsection \ref{subsec: dust evolution}, the relative velocity between the dust and the gas changes in the vicinity of the warp, as a result of the change in the gas surface density. Consequently, aeolian-erosion (e.g. \citealp{Erosion1})
 would be suppressed, although it might still function if the disk is turbulent.
Moreover, streaming-instability \citep{YoudinGoodman2005} could be affected by the different structures of the disk and the new redistribution of solids in the vicinity of the warp and in further areas that are affected as well. Breaking the disk into several small disks might limit the size abundance and by that increase the efficiency of streaming-instability \citep{MartinLubow2022}.
Misalignments could lead also to a change in the relative velocities between objects in the disk, such that destructive collisions and collisional growth might be affected as well. Furtheremore, the size distribution and mass segregation of objects in the disk could be modified due to the warping.

\section{Caveats }\label{sec:caveats}

Here we will briefly discuss potential caveats of our model:

\begin{itemize}
    \item The formation of dust traps could lead to a backreaction in the dust evolution/scale-height, which in turn will strengthen the trapping of dust particles and the efficiency of the  planet formation and one should take into account the combined evolution of the backreaction and the formation of dust ring due to the perturbation
(see discussion on these topics in \citealp{Gonzalez2017,Longarini2021}).

\item The warping could modify also the pressure gradient, which will add another correction to the radial drift. This is not taken into account in our current derivation. 

\item We neglected the vertical motion of the objects, but in general the warping will change the behavior in this direction as well. 
\end{itemize}

\section{Summary}\label{sec:summary}

In this paper, we discussed the radial drift in warped disks. We derived analytically the expression for dust radial drift in these disks and demonstrated for some examples of warped disks. Radial drift in warped disks is essentially different from the one in flat disk and might have important consequences on the nature of planet formation in these disks. Not only that warped disks could give rise to dust traps in which the radial drift halts, but they also modify the radial drift outside the traps.

While most of the studies of dust traps focus on pressure maxima, in distorted disks in general and warped disks especially, there could be another type of dust traps, which rises from the different precession rates of the gas and the dust.
These traps could interact also with other growth and destruction mechanisms of planets, and should be taken into account as an integral part of the planet formation theory in distorted disks. The characteristics of the perturbed profile are derived from the origin of the perturbation, which could be either transient (e.g. for flybys) or long-lasting (e.g. circumbinary disks). 

The traditional planet formation research focused on coplanar single star disks, but there is a wealth of distorted disks, which should be included in planet formation theories, especially due to the piling-up evidence of these disks. Not only the radial drift changes, but also further steps in the evolution of planets and planetesimals, and the physical picture is far from being complete.

%\section*{ACKNOWLEDGMENTS}  
%I would like to thank Hagai B. Perets and Yonadav Barry Ginat for fruitful discussions. 
%MR acknowledges the generous support of Azrieli fellowship. 
%%%%%%%%%%%%%%%%%%%%%%%%%%%%%%%%%%%%%%%%%%%%%%%%%

%%%%%%%%%%%%%%%%%%%% REFERENCES %%%%%%%%%%%%%%%%%%

% The best way to enter references is to use BibTeX:

\bibliographystyle{aasjournal}
\bibliography{example} % if your bibtex 
%%%%%%%%%%%%%%%%%%%%%%%%%%%%%%%%%%%%%%%%%%%%%%%%%%
%%%%%%%%%%%%%%%%%%%%%%%%%%%%%%%%%%%%%%%%%%%%%%%%%%

\appendix

\section{Radial drift in a warped disk -- full derivation}\label{app: full derivation}

Here we derive the radial drift for a warped disk, generalizing the derivation for a flat disk in \cite{Nakagawa1986}.

The equations of motion for the dust and gas particles are given by (e.g. \citealp{Nakagawa1986})

\begin{align}
&\frac{d\textbf u_d}{dt} =-A \rho_g(\textbf{u}_d-\textbf{u}_g)-\frac{GM_\star}{r_d^3}\textbf{r}_d, \\
&\frac{d\textbf u_g}{dt} =-A \rho_g(\textbf{u}_g-\textbf{u}_d)-\frac{GM_\star}{r_g^3}\textbf{r}_g-\frac{\nabla P_g}{\rho_g}
\end{align}

\noindent
where $A$ is the gas drag coefficient and from now parameters related to the dust will be noted with subindex $d$ and parameters related to the gas with subindex $g$.
To obtain the gas and dust velocities in a warped disk, we start from these velocities in a flat disk, and then rotate it by an angle $\beta=\beta_g=\beta_d$ around the y-axis and $\gamma$ around the $z$ axis, where $\gamma_g=\Omega_p\textbf{t}$ for the gas and $\gamma_d=\Omega_{\rm ext}\textbf{t}$ for the dust. 
To generalize the derivation specified in \cite{Nakagawa1986} for a warped disk, we will introduce the tilt and twist angle $\beta$ and $\gamma_i$ correspondingly and rotate the general flat velocity fields, when $u_i'$ refer to the velocity components in the flat system.

\begin{align}\label{eq:u decomposition}
\textbf{u}_d=\begin{pmatrix}
u'_{d,r}\left(\cos\phi\cos \beta\cos \gamma_d-\sin \gamma_d\sin \phi\right)-u'_{d,\phi}\left(\sin \phi \cos \beta \cos \gamma_d+\cos \phi \sin \gamma_d\right) \\
u'_{d,r}\left(\cos \phi \sin \gamma_d \cos \beta +\cos \gamma_d\sin \phi\right)+u'_{d,\phi}\left(\cos \phi\cos \gamma_d-\sin \phi \sin \gamma_d\cos \beta\right)\\
-u'_{d,r}\sin \beta\cos \phi +u'_{d,\phi}\sin \phi \sin \beta
\end{pmatrix},\\~\\
\textbf{u}_g=\begin{pmatrix}
u'_{g,r}\left(\cos\phi\cos \beta\cos \gamma_g-\sin \gamma_g\sin \phi\right)-u'_{g,\phi}\left(\sin \phi \cos \beta \cos \gamma_g+\cos \phi \sin \gamma_g\right) \\
u'_{g,r}\left(\cos \phi \sin \gamma_g \cos \beta +\cos \gamma_g\sin \phi\right)+u'_{g,\phi}\left(\cos \phi\cos \gamma_g-\sin \phi \sin \gamma_g\cos \beta\right)\\
-u'_{g,r}\sin \beta\cos \phi +u'_{g,\phi}\sin \phi \sin \beta
\end{pmatrix}
\end{align}

The equations of motion in the flat system of coordinates are then given by, when we neglect the vertical dynamics, i.e. the $\hat z$ components of the equations 
and define $\textbf{v}_d=\textbf{u}_d-\textbf v_{\rm Kep}, \ \textbf{v}_g=\textbf{u}_g-\textbf v_{\rm Kep}$ are given by 

\begin{align}
&\underline{d,\hat r':} \ \frac{\partial \textbf{v}_{d}}{\partial t}\cdot \hat r'= \left[-A \rho_g(\textbf{v}_d-\textbf{v}_g)-\frac{GM_\star}{r_d^3}\textbf{r}_d-\frac{\partial \textbf v_{\rm Kep}}{\partial t}\right]
\cdot \hat r',\\
&\underline{d,\hat \phi':} \ \frac{\partial \textbf v_{\rm d}}{\partial t}\cdot \hat \phi'= \left[-A \rho_g(\textbf{v}_d-\textbf{v}_g)-\frac{GM_\star}{r_d^3}\textbf{r}_d-\frac{\partial \textbf v_{\rm Kep}}{\partial t}\right]
\cdot \hat \phi', \\
&\underline{g,\hat r':} \ \frac{\partial \textbf{v}_{g}}{\partial t}\cdot \hat r'= \left[-A \rho_g(\textbf{v}_g-\textbf{v}_d)-\frac{GM_\star}{r_g^3}\textbf{r}_g-\frac{\partial \textbf v_{\rm Kep}}{\partial t}-\frac{\nabla P_g}{\rho_g}\right]
\cdot \hat r', \\
&\underline{g,\hat \phi':} \ \frac{\partial \textbf{v}_{g}}{\partial t}\cdot \hat \phi'= \left[-A \rho_g(\textbf{v}_g-\textbf{v}_d)-\frac{GM_\star}{r_g^3}\textbf{r}_g-\frac{\partial \textbf v_{\rm Kep}}{\partial t}\right]
\cdot \hat \phi'
\end{align}

Using the velocity decomposition in terms of the velocity components in the flat coordinate system specified in eq. \ref{eq:u decomposition}, 

\footnotesize
\begin{align}
\nonumber
&\underline{d,\hat r':} \\
&B_{r}(\gamma_d)\frac{\partial v'_{d,r}}{\partial t}+\ B_{\phi}(\gamma_d)\frac{\partial v'_{d,\phi}}{\partial t}\approx -A\rho_g\left[C_{r}(\gamma_d)v_{d,r}'+C_{\phi}(\gamma_d)v'_{d,\phi}-C_r(\gamma_g)v'_{g,r}-C_{\phi}(\gamma_g)v'_{g,\phi}\right]+Q(\gamma_d)2\Omega_K v'_{d,\phi}-S(\gamma_d)\frac{1}{2}\Omega_Kv_{d,r}', \\
&\underline{d,\hat \phi':} \nonumber
\\
&F_{r}(\gamma_d)\frac{\partial v'_{d,r}}{\partial t}+F_\phi(\gamma_d)\frac{\partial v'_{d,\phi}}{\partial t}\approx -A\rho_g\left[G_{r}(\gamma_d)v_{d,r}'+G_{\phi}(\gamma_d)v'_{d,\phi}-G_r(\gamma_g)v'_{g,r}-G_{\phi}(\gamma_g)v'_{g,\phi}\right]+U(\gamma_d)2\Omega_Kv'_{d,\phi}-V(\gamma_d)\frac{1}{2}\Omega_K v'_{d,r},\\
~
\\
&\underline{g,\hat r':} \nonumber
\\
&B_{r}(\gamma_g)\frac{\partial v'_{g,r}}{\partial t}+\ B_{\phi}(\gamma_g)\frac{\partial v'_{g,\phi}}{\partial t}\approx -A\rho_d\left[C_{r}(\gamma_g)v_{g,r}'+C_{\phi}(\gamma_g)v'_{g,\phi}-C_r(\gamma_d)v'_{d,r}-C_{\phi}(\gamma_d)v'_{d,\phi}\right]+Q(\gamma_g)2\Omega_K v'_{g,\phi}-S(\gamma_g)\frac{1}{2}\Omega_Kv_{g,r}'-\frac{1}{\rho_g}\frac{\partial P_g}{\partial r'}, \\
&\underline{g,\hat \phi':} \nonumber
\\ &F_{r}(\gamma_g)\frac{\partial v'_{g,r}}{\partial t}+F_\phi(\gamma_g)\frac{\partial v'_{g,\phi}}{\partial t}\approx -A\rho_g\left[G_{r}(\gamma_g)v_{g,r}'+G_{\phi}(\gamma_g)v'_{g,\phi}-G_r(\gamma_d)v'_{d,r}-G_{\phi}(\gamma_d)v'_{d,\phi}\right]+U(\gamma_g)2\Omega_Kv_{d,\phi}'-\textbf{V}(\gamma_g)\frac{1}{2}\Omega_K v'_{g,r}
\end{align}

\normalsize
When the coefficients are calculated by projecting the velocity vectors on the 'flat' coordinate system\textbf{ and extracting the coefficients: $\hat r\cdot \hat u_i:=C_r(\gamma_i) u'_{i,r}+C_{\phi}(\gamma_i)u'_{i,\phi}$ and $\hat \phi \cdot \hat u_i=G_r(\gamma_i)u'_{i,r}+G_{\phi}(\gamma_i)u'_{i,\phi}$ which yields}

\begin{align}
&C_r(\gamma_i) = \cos \beta\cos \phi \cos \left(\phi-\gamma_i\right)+\sin \phi\sin\left(\phi-\gamma_i\right),\\
&C_\phi(\gamma_i)=\cos \gamma_i \sin^2 \frac{\beta}{\textbf 2}\sin(2\phi)-\sin \gamma_i\left(\cos^2 \phi+\cos \beta \sin^2 \phi\right),\\
&G_r(\gamma_i)= \cos(\phi-\gamma_i)\sin \phi-\cos\beta \cos \phi \sin(\phi-\gamma_i),\\
&G_\phi(\gamma_i)=\cos(\phi-\gamma_i)\cos\phi+\cos \beta \sin \phi \sin(\phi-\gamma_i)
%,\\
%&B_r(\gamma_i)\approx C_r(\gamma_i), \ B_\phi(\gamma_i)\approx C_\phi(\gamma_i),\\
%&F_r(\gamma_i)\approx G_r(\gamma_i), \ F_\phi(\gamma_i)\approx G_\phi(\gamma_i), \\
%&Q(\gamma_i)\approx 1-C_r(\gamma_i), \ S(\gamma_i)\approx C_\phi(\gamma_i), \\
%&U(\gamma_i)\approx G_r(\gamma_i), \ V(\gamma_i)=-G_{\phi}(\gamma_i)+1
\end{align}

\noindent
$B_i,F_i,Q_i,S_i,U_i,V_i$ could be calculated in the same procedure, but as we focus on a specific case of steady-state solutions, $B_i,F_i$ won't play a role. We will consider $Q_i=V_i=1, \ S_i=U_i=0$, similarly to the calculation in a flat disk, as a simplifying assumption. To calculate the effective change in radial drift, one could consider the orbit-averaged coefficients. 
\begin{align}
&\bar C_r(\gamma_i)=\frac{1}{2\pi}\sin\left(2\pi \frac{\Omega_p}{\Omega_{\rm bin}}\right)\frac{\Omega_{\rm bin}}{\Omega_p}\left(\frac{\Omega_p}{\Omega_{\rm bin}}-2\right)^{-1}\left[\cos \beta\left(\frac{\Omega_p}{\Omega_{\rm bin}}-1\right)-1\right], \\
&\bar C_\phi(\gamma_i)= -\frac{1}{\pi}\left(\frac{\Omega_p}{\Omega_{\rm bin}}-2\right)^{-1}\sin^2 \left(\pi\frac{\Omega_p}{\Omega_{\rm bin}}\right)
\left[-\frac{\Omega_{p}}{\Omega_{\rm bin}}+1+\cos \beta \right],\\
%\left[2\sin^2\frac{\beta}{2}+\left(\frac{\Omega_p}{\Omega_{\rm bin}}\right)^2-2-2\cos \beta \right], \\
&\bar G_r(\gamma_i)=\frac{1}{\pi}\sin^2 \left(\pi \frac{\Omega_p}{\Omega_{\rm bin}}\right)\left(\frac{\Omega_p}{\Omega_{\rm bin}}-2\right)^{-1}\left[\cos \beta\left(\frac{\Omega_p}{\Omega_{\rm bin}}-1\right)-1\right],\\
&\bar G_\phi(\gamma_i)= \frac{1}{2\pi}\sin\left(2\pi \frac{\Omega_p}{\Omega_{\rm bin}}\right)\frac{\Omega_{\rm bin}}{\Omega_p}\left(\frac{\Omega_p}{\Omega_{\rm bin}}-2\right)^{-1}\left[\frac{\Omega_p}{\Omega_{\rm bin}}-1-\cos \beta\right]
\end{align}
%The dependence on $\phi$ in the coefficients could be averaged over the disk, such the averaged coefficients are

%\begin{align}
%&\bar C_r(\gamma_i)= \frac{1}{2}\cos \gamma_i\left(1+\cos \beta\right) , \\
%&\bar C_\phi(\gamma_i)= -\frac{1}{2}\sin \gamma_i \left(1+\cos \beta\right), \\ 
%&\bar G_r(\gamma_i)= \frac{1}{2}\sin \gamma_i\left(1+\cos \beta\right), \\
%&\bar G_\phi(\gamma_i)= \frac{1}{2}\cos \gamma_i\left(1+\cos \beta\right)
%\end{align}
Then, a full generalized analytical solution for the dust radial drift at a steady-state (i.e. the partial derivatives equal to zero) could be derived (although complicated). Hence, we will examine some simplifying assumptions: $\gamma_g=\gamma_d$ i.e $C_r:=C_r(\gamma_d)=C_r(\gamma_g), \ C_\phi:=C_\phi(\gamma_d)=C_\phi(\gamma_g)$ and $G_r:=G_r(\gamma_d)=G_r(\gamma_g), \ G_\phi:=G_\phi(\gamma_d)=G_\phi(\gamma_g), Q(\gamma_d)=Q(\gamma_g)=1, \ S(\gamma_g)=S(\gamma_d)=0, \ U(\gamma_d)=U(\gamma_g)=0, \ V(\gamma_d)=V(\gamma_g)=0$ --- which takes place (but not only) at the co-precession radius, and neglecting the corrections to the derivatives on the left side, the radial drift in the flat system is 

\noindent

\scriptsize
\begin{align}
&v_{d,r}=
\frac{A^3 C_r \Omega_K^2 \frac{1}{\rho_g}\frac{\partial P_g}{\partial r} \rho_g^3 \left(-G_\phi^2-2G_\phi G_r-G_r^2\right)}{0.5 A^3 \Omega^2
   \rho_g^2  C_r  (-2 G_\phi \rho_d-2 G_\phi \rho_g-2 G_r \rho_d-2 G_r
   \rho_g) (A C_r ( G_\phi \rho_d+G_\phi \rho_g+G_r \rho_d+ G_r \rho_g)-0.5
   C_\phi \Omega_K)+A^2 C_r \Omega_K^4 \rho_g^2 (-G_\phi-G_r)}= \\ \nonumber
  & = 
  -2\eta \Omega_K v_K \frac{A^3 C_r \Omega_K^2  \rho_g^3 \left(-G_\phi^2-2G_\phi G_r-G_r^2\right)}{0.5 A^3 \Omega^2
   \rho_g^2  C_r  (-2 G_\phi \rho_d-2 G_\phi \rho_g-2 G_r \rho_d-2 G_r
   \rho_g) (A C_r ( G_\phi \rho_d+G_\phi \rho_g+G_r \rho_d+ G_r \rho_g)-0.5
   C_\phi \Omega_K)+A^2 C_r \Omega_K^4 \rho_g^2 (-G_\phi-G_r)}= 
   \\
   \nonumber
  & = -2\eta \Omega_K v_K \frac{-A^3C_r\Omega_K^2\rho_g^3(G_\phi+G_r)^2}{-A^2\Omega_K^2\rho_g^2C_rD(G_\phi+G_r)\left[C_r(G_r+G_\phi)D-0.5C_\phi \Omega_K\right]-A^2C_r\Omega_K^4\rho_g^2(G_\phi+G_r)}= \\
  \nonumber
  &=
   -2\eta \Omega_K v_K D\frac{\rho_g}{\rho_g+\rho_d} \frac{G_\phi+G_r}{D\left[C_r(G_r+G_\phi)D-0.5C_\phi \Omega_K\right]+\Omega_K^2}
\end{align}

\normalsize
\noindent
where $D=A(\rho_g+\rho_d)$ and the last reduction is not trivial and holds only when $G_r+G_\phi, C_r\neq 0$.

\section{Limiting cases}
Here we will examine our results in some limiting cases.
\\
\subsection{The standard radial drift limit ($\beta,\gamma_i\to 0$)}
For this case, in which the tilt angle and the precession are negligible, 
\begin{align}
C_r, G_\phi\to 1, \ 
C_\phi, G_r \to 0  
\end{align}
\noindent
The equations then become
\begin{align}
&\underline{d,\hat r':} \ 0\approx -A\rho_g \left(v_{d,r}'-v_{g,r}'\right)+2\Omega_K v_{d,\phi}'\\
&\underline{d,\hat \phi':} \ 0\approx -A\rho_g \left(v_{d,\phi}'-v_{g,\phi}'\right)-\frac{1}{2}\Omega_K v_{d,r}'\\
&\underline{g,\hat r':} \ 0\approx -A\rho_g \left(v_{g,r}'-v_{d,r}'\right)+2\Omega_K v_{d,\phi}'-\frac{1}{\rho_g}\frac{\partial P_g}{\partial r'}, \\
&\underline{g,\hat \phi':} \ 0\approx -A\rho_g \left(v_{g,r}'-v_{d,r}'\right)-\frac{1}{2}\Omega_K v_{g,r}'
\end{align}
\noindent
which are the standard equations for radial drift in a flat disk (e.g. \citealp{Nakagawa1986}), i.e. $\xi\to 1$.
\subsection{Small warping, negligible precession ($0<\beta\ll 1, \ \gamma_i\to 0$)}
In the limit of small warping and negligible precession, the coefficients are given by 
\begin{align}
&C_r\approx 1-\frac{\beta^2}{2}\cos^2\phi, \\
& C_\phi\approx \frac{\beta^2}{2}\sin(2\phi), \\
& G_r\approx \frac{\beta^2}{2}\cos \phi \sin \phi, \\
& G_\phi \approx 1-\frac{\beta^2}{2}\sin^2 \phi
\end{align}
\noindent
and the orbit-averaged coefficients are 
\begin{align}
&\bar C_r=\bar G_\phi\approx 1-\frac{\beta^2}{4}, \\
&\bar C_\phi =\bar G_r \approx 0
\end{align}
\noindent
Then, the correction factor to the radial drift $\xi$ is given by 
\begin{align}
\xi\approx \frac{1}{1-\frac{\beta^2}{4}}\frac{D^2+\Omega_K^2}{D^2+\left(\frac{\Omega_K}{1-0.25\beta^2}\right)^2}
\end{align}

\end{document}